\documentclass[aps,twocolumn,secnumarabic,nobalancelastpage,amsmath,amssymb,
nofootinbib]{revtex4}

\usepackage{pdfpages}
\usepackage{graphics}     
\usepackage{graphicx}      
\usepackage{longtable}     
\usepackage{url}          
\usepackage{bm}          
\usepackage{mathtools,amsmath,amssymb,lipsum,xcolor}

\begin{document}
\title{Time Glass: A Fractional Calculus Approach }
\author         {R.~C. Verstraten$^{1}$, R.~F. Ozela$^{1,2}$, C. Morais Smith$^{1}$}
\affiliation    {$^{1}$\mbox{Institute for Theoretical Physics, Utrecht University, Princetonplein 5, 3584CC Utrecht, The Netherlands} \\
$^{2}$Faculdade de F\'{\i}sica, Universidade Federal do Par\'a,  66075-110 Bel\'em, PA,  Brazil}
\date{\today}

\begin{abstract}
Out of equilibrium states in glasses and crystals have been a major topic of research in condensed-matter physics for many years, and the idea of time crystals has triggered a flurry of new research. Here, we provide the first description for the recently conjectured Time Glasses using fractional calculus methods. An exactly solvable effective theory is introduced, with a continuous parameter describing the transition from liquid through normal glass, Time Glass, into the Gardner phase. The phenomenological description with a fractional Langevin equation is connected to a microscopic model of a particle in a sub-Ohmic bath in the framework of a generalized Caldeira-Leggett model.
\end{abstract}



\maketitle
\textbf{Introduction.} 
The concept of time crystals, i.e., materials with an emergent periodicty in time~\cite{shapere2012classical,wilczek2012quantum} has attracted much attention recently~\cite{khemani2019brief}. After no-go theorems were proven for several systems~\cite{bruno2013impossibility,watanabe2015prl}, it was understood that time crystals may actually occur in open~\cite{iemini2018boundary,buvca2019non}, driven~\cite{khemani2016phase}, and long-range interaction~\cite{valerii2019longrange} systems, and many  experimental and theoretical activities followed this novel idea. By now, time crystals have been conceived and observed~\cite{zhang2017observation, rovny2018observation, li2020discrete, yao2020classical,medenjak2019isolated}. 
Recently, a novel system was conjectured, a Time Glass, which would have periodic intermediate glass states~\cite{wilczek2019timecrystals}. It was claimed that they should appear in static many-body localized systems exhibiting incommensurate local frequencies and no long-range spatiotemporal order~\cite{khemani2019brief}.

The microscopic states of glasses have been puzzling researchers for many years~\cite{debenedetti2001supercooled,lubchenko2015theory,schoenholz2016structural,berthier2017active,ninarello2017models,liao2019hierarchical}. Microscopically, glasses look like liquids because their molecules do not show any kind of structural order. These amorphous materials exhibit completely different phase transitions in comparison to ordered solids and often they do not even have a thermodynamic ground state. Hence, equilibrium physics cannot be used for their description~\cite{cugliandolo1994out}. A common understanding is that they correspond to the occupation of a set of metastable states in the free-energy landscape. 

\begin{figure}[h]
	\centering
	\includegraphics[width=0.9\columnwidth]{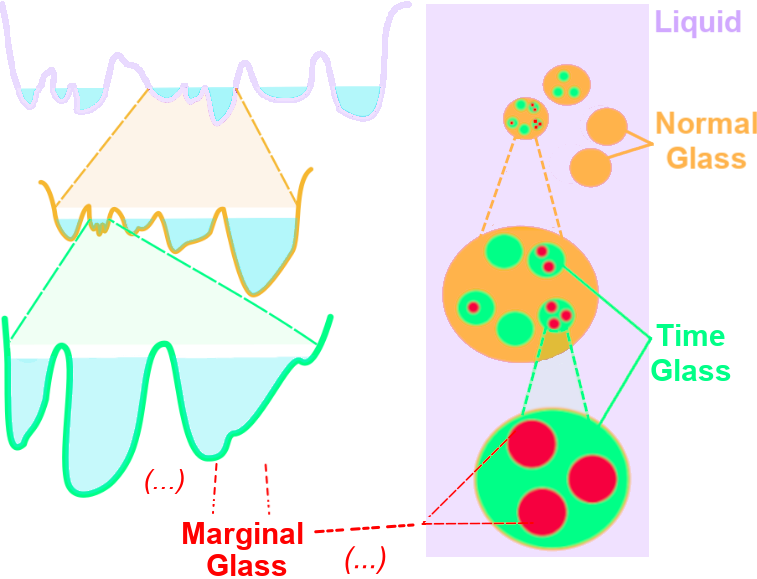}
	\caption{Left: Sketch of a  free energy landscape with fractal-like metastable minima. Right: A hierarchy of cages inside cages. The energy landscape (purple) has four local minima, corresponding to the total number of large orange cages. Zooming in on one of these minima, we see five smaller local minima (green cages) inside this orange cage.  Moreover, inside one green cage there are three red cages. Depending on the number of nested cages we find 4 differing phases: Liquid (no cage), normal glass (1 cage), Time Glass (finite number of cages), and marginal glass (infinite number of cages).}
	\label{fig:zooming}
\end{figure}

One promising approach to describe the glass transition is the random first-order transition theory~\cite{biroli2008thermodynamic}. It describes a hard-sphere model, in which the spheres can get caged by their neighbors, thus restricting their movement. This leads to a mean square displacement (MSD) that corresponds to a free particle at small time scales, until it hits the cage size (see Fig.~\ref{fig:zooming}), where the MSD saturates, indicating a glassy state~\cite{berthier2011theoretical,flores2012msd} (for a discussion on glass states, see SM). In the free energy landscape, this is the moment at which the particle has explored the entire basin. This theory, however, can only be solved exactly in infinite dimension, with a questionable connection to its finite-dimensional counterpart~\cite{biroli2012random}. 

More recently, a richer phase diagram was proposed for glasses, including the so-called Gardner phase~\cite{berthier2019gardner,li2021determining}. Although the Gardner phase was first discovered in the description of spin glasses as a solution that breaks one replica symmetry~\cite{gardner1985spin}, it was later understood to occur in many materials~\cite{seguin2016experimental}. This marginal glass phase has a free energy in which basins transform into metabasins~\cite{charbonneau2017glass}, and is known to have a fractal structure~\cite{charbonneau2014fractal} (see Fig.~\ref{fig:zooming}).  In the Gardner phase, there is a hierarchy of cages inside cages~\cite{charbonneau2015numerical}, reminiscent of the fractal structure in the energy landscape. Therefore, as time goes by, the system explores larger cages, thus triggering an infinite staircase-like behavior of the MSD. 

Here, we propose a unified picture where liquid, glass, Time Glass, and the Gardner phase can be understood as a (sub-)diffusive Brownian motion described by a fractional Langevin equation with white noise. Previously, the fractional Langevin equation had been studied for colored noise~\cite{metzler2000random,hilfer2000applications,lutz2001langevin,barkai2001fractional,burov2008critical,deng2009ergodic,pottier2003aging,metzler2014anomalous,jeon2010fractional,jeon2013anomalous,lizana2010foundation,vojta2019probability}, but the features described here occur exclusively for the case of white noise. At very low temperatures, the equipartition theorem breaks down and a white-noise fractional Langevin equation arises as the semiclassical description of a particle coupled to a two-level systems (TLS) bath in the subohmic regime. Our unified procedure has a single varying generic $s$-derivative friction, with $s$ integer or fractional, to describe the different states of matter. Moreover, we show that for $0 < s \lesssim 0.1$, a \textit{Time Glass} emerges, with a periodicity of $\pi (M/\eta)^{1/(2-s)}$, where $M$ is the mass of the ``Brownian'' particle and $\eta$ is a type of viscoelasticity. Within our mean-field description, the Time Glass shows an emergent frequency and periodically passes through many metastable states before it gets to a final frozen glassy state. Therefore, our work sets a mathematical framework for the description and realization of a Time Glass, thus furthering our understanding of ``time materials''.

\textbf{Fractional Calculus.} 
Fractional Differential Equations have been used in physics, engineering, material science, control systems, protein folding~\cite{metzler2000random}, and more~\cite{sun2018new},  but there are still many new opportunities to be explored~\cite{hilfer2000applications}.
Different definitions were proposed by Riemann-Liouville, Caputo, Weyl, \textit{etc}. (see SM for a short historical overview). From a mathematical perspective, there is still a discussion on which of the various fractional derivative definitions~\cite{longlist} should be used for each kind of problem. Concerning the  Caputo definition, the idea is to rewrite a repeated 
integral into a generalizable form. As factorials often appear in conventional integrals, one uses the Gamma function as their non-integer generalization. The Caputo derivative is then given by taking an integer derivative before doing a fractional integral;
\begin{equation}
\mathbf{D}_{t}^\alpha f(t)	= \frac{1}{\Gamma(n-\alpha)} \int_0^t (t-\tau)^{n-\alpha-1} f^{(n)}(\tau) \text{ d}\tau,
\end{equation}
where $n$ is an integer such that $n-1\le\alpha<n$. Since the Caputo definition is non-local, we have chosen the left-handed definition for the boundary of the integral to retain causality, once we apply this time derivative to our system. One benefit of the Caputo definition compared to other definitions is that we can keep integer-order boundary conditions; however, continuity in the order is lost on the integers. When this non-integer derivative is applied to an exponential, we find the Mittag-Leffler function, defined by
\begin{equation}
E_{\alpha,\beta}(t)= \sum_{k=0}^\infty \frac{t^k}{\Gamma(\alpha k+\beta)},
\end{equation}
which is a generalized exponential that appears regularly in solutions of fractional differential equations. For different parameters, this function can show many features related to exponentials, such as damped oscillations and exponential-like growth.

\textbf{Fractional Langevin Equation.} 
The fractional Langevin equation was recently used to describe a system exhibiting L\'evy flights~\cite{vojta2019probability}. L\'evy flights are often used for modeling the spreading of viruses, as they include a description of the long (and fast) journeys that people make by plane, as well as a more local random motion. Brownian motion only has one typical time and length scale associated to it, while L\'evy flights have many different time and length scales~\cite{metzler2000random,eliazar2013motions}. The equipartition theorem implies a relation between the fluctuation and dissipation terms~\cite{kubo1966fluctuation}; hence, previous references used either colored noise with the fractional Langevin equation~\cite{metzler2000random,hilfer2000applications,lutz2001langevin,barkai2001fractional,burov2008critical,lizana2010foundation,vojta2019probability} or a white noise associated to a fractional kinetic term and normal friction~\cite{kobelev2000fractional} which, after fractional integration, is equivalent to the fractional Langevin equation with colored noise. However, none of these models exhibit the plateaus in the MSD characterizing a glassy behavior, which we will describe below. The features that we will discuss here are inherent to a fractional Langevin equation with white noise, and occur only at low temperatures, when the equipartition theorem breaks down. Later, in the \textit{Microscopic Model} Section, we present a microscopic description corresponding to a physical realization of our phenomenological model.


Let us start by considering the \textit{fractional Langevin equation}
\begin{equation}
M\frac{\text{d}^2 x(t)}{\text{d}t^2} + \eta \mathbf{D}_{t}^s x(t) = f(t),
\end{equation}
with $f(t)$ a white-noise force with average $\langle f(t)\rangle= 0$ and correlation $\langle f(t)f(t')\rangle=K\delta(t-t')$, where $K=2 \sin\left(\frac{\pi s}{2}\right)\eta  t_s^{1-s} k_B T$ and $t_s= \left(M/\eta\right)^{\frac{1}{2-s}}$ is the time scale of the system. Introducing these fractional derivatives as friction, scaling with the $s^{th}$ order derivative, yields a model for both sub-diffusion $(s < 1)$ and super-diffusion $(s > 1)$, in addition to the usual Langevin Equation, which is retrieved when $s=1$. The motivation for this change of friction compared to the Langevin equation is firstly to introduce a general non-local operator that knows about the history of a system and allows one to study non-Markovian processes. The choice to take fractional derivatives can then be illustrated with a thought experiment: Suppose a particle is moving at a constant speed for a certain time. Then, ordinary friction is constant in time, while friction of this fractional form scales as $t^{1-s}$. This means that for $s>1$ the friction will fall off quickly, allowing for L\'evy flights, while for $s<1$ the friction will increase in time, thus reducing the probability for large jumps~\cite{bovet2015introduction}. 

	\begin{figure*}[t]
	\hspace*{-1cm}
	\includegraphics[width=\textwidth]{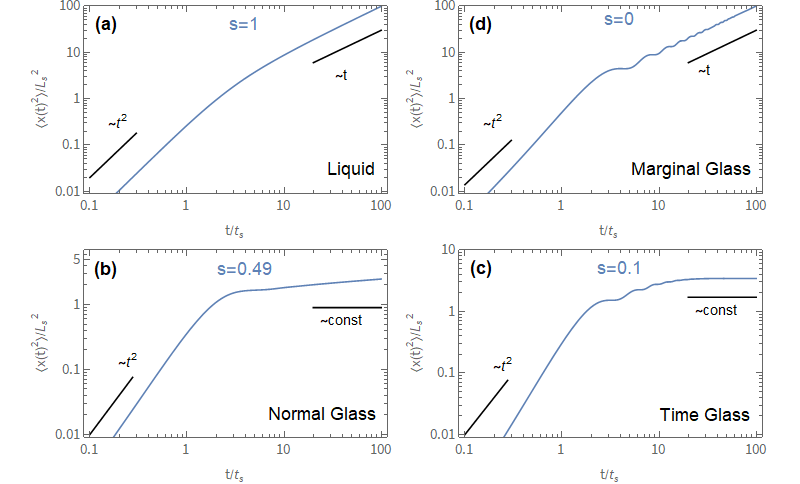}
	\caption{The exact MSD for different values of $s$, regularized by the typical time and length scales $t_s$ and $L_s$. The analytical asymptotes for $t\to 0$ and $t\to\infty$ are drawn next to the plots. The delocalized states are shown at the top and the localized states at the bottom, while the labeling order follows the decreasing value of $s$.  \textbf{(a)} A liquid, described by regular Brownian motion. \textbf{(b)} A normal glass, exhibiting ballistic motion at short times and localization at longer times.  \textbf{(c)} A Time Glass, displaying ballistic motion at short times, but also an intermediate regime with a set of small plateaus, before the long-time plateau sets in. \textbf{(d)} A marginal glass, characterized by a driven particle in a quadratic potential. Here, the friction term vanishes and an infinite collection of plateaus appear. The overall slope is however finite, reminiscent of a liquid behavior. }
	\label{fig:msd}
\end{figure*}

 The strength of this theory is that it is exactly solvable, which makes possible the calculation of statistical properties, such as the MSD. For $0\le s<1$, the MSD is given by (see SM for details)
\begin{align}
\langle x(t)^2 \rangle &= \frac{K}{M^2} \int_0^t  \left[\tau E_{2-s,2}\left(-\frac{\eta}{M}     \tau^{2-s}\right) \right]^2\;d\tau \nonumber\\
&\qquad+\left[ v_0 L_s \frac{t}{t_s} E_{2-s,2}\left( -\frac{\eta}{M} t^{2-s} \right) \right]^2 ,
\end{align}
where $v_0 = \langle x'(0)\rangle t_s/L_s$ is dimensionless.
The short-time expansion for $t\ll t_s$ yields a ballistic behavior,
\begin{equation}
  \langle x(t)^2 \rangle = (v_0 L_s)^2 \left(\frac{t}{t_s}\right)^2,
\end{equation} 
while the long-time expansion is logarithmic for $s=0.5$, and otherwise given by
\begin{equation}
 \langle x(t)^2 \rangle= \Delta_s^2 + \frac{K t^{2s-1}}{(2s-1)\eta^2\Gamma(s)^2},
\end{equation}
where, for $s<1/2$, the exponent of the second term in the MSD becomes negative and, therefore, the MSD converges to the typical final cage size $\Delta_s^2$ for $t\to \infty$.

 In Fig.~\ref{fig:msd}, the MSD has been plotted for several values of $s$ from one to zero. We introduced the typical length scale of the system, which can be found by dimensional analysis to be $L_s= \sqrt{K t_s^3 /M^2}$. The MSD shows ballistic short-time behavior in all cases. For $s=1$, we retrieve the conventional Langevin equation, which describes Brownian motion. The MSD also shows a crossover from a ballistic ($\sim t^2$) to a linear dependence in time, characteristic of a liquid [Fig.~\ref{fig:msd}(a)]. For $s \lesssim 0.5$, instead, the MSD saturates at large times, thus describing a glass [Fig.~\ref{fig:msd}(b)]. We find that a particularly interesting regime is provided by small values of $s$, in the interval $0<s\lesssim0.1$. In this case, a sequence of small metastable plateaus characterizes a finite-depth fractal glass phase, before the conventional glass regime is reached at larger times [Fig.~\ref{fig:msd}(c)]. For $s=0$, the ``marginal glass'' phase, proposed by Gardner, is realized,  with an infinite number of metastable plateaus and finite average slope ($\sim t$), typical of liquids. This is an asymptotic phase, in which the fractal glass acquires infinite depth [Fig.~\ref{fig:msd}(d)].

  	\begin{figure}[t!]
	\centering
	\includegraphics[width=\columnwidth]{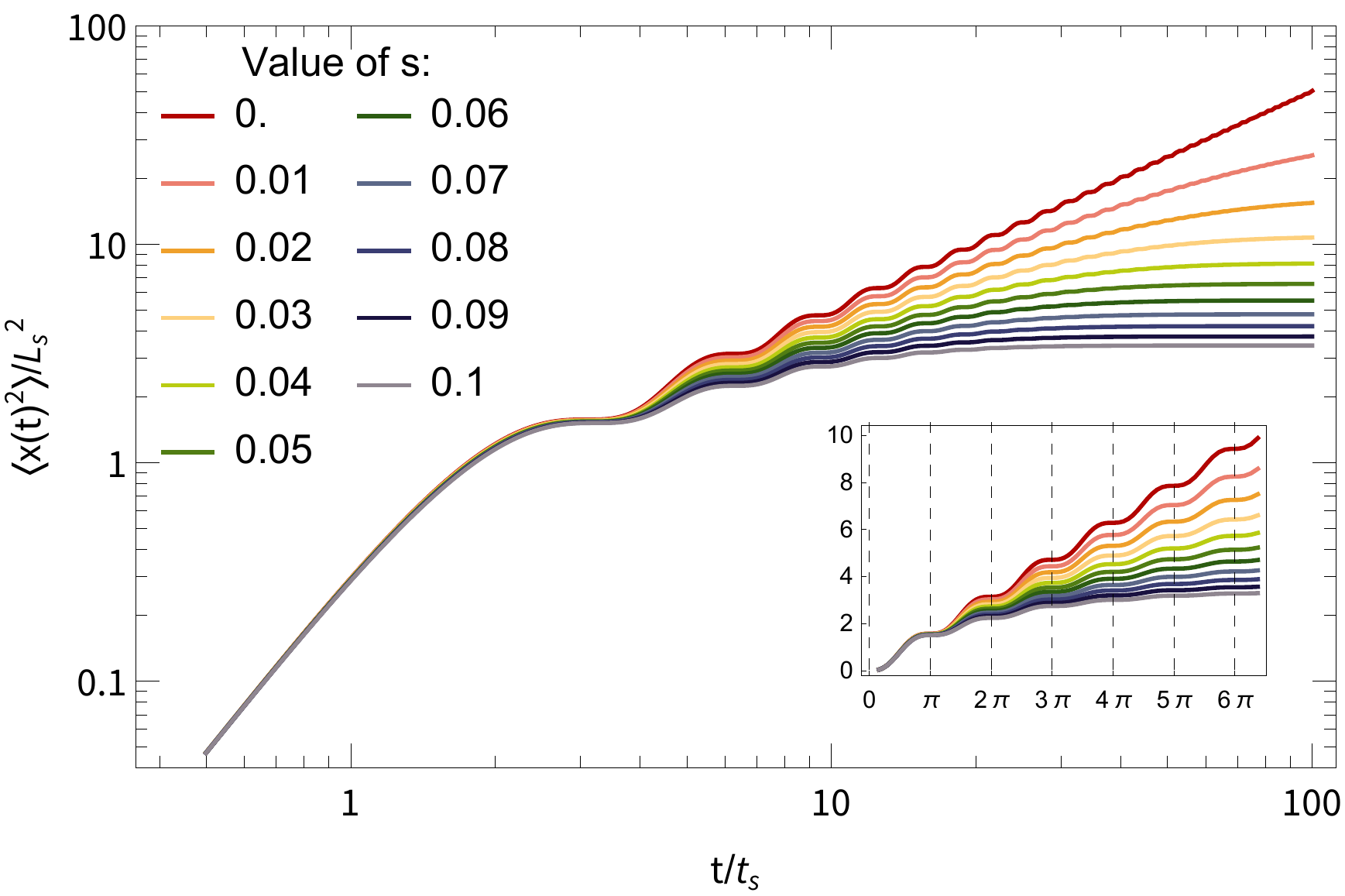}
	\caption{The MSD for several small values of $s$, regularized by the typical time and length scales $t_s$ and $L_s$. The inset shows a linear scale plot to highlight the periodicity in the anomalous glass phase, characterizing a Time Glass.}
	\label{fig:msdsmallbeta}
\end{figure}

\textbf{Time Glass.} 
Recently conjectured~\cite{wilczek2019timecrystals}, \textit{Time Glass} is a phase in which the MSD periodically repeats a glass-like plateau in an overall glassy phase (see SM for a comparison between definitions of Time Glass).
Now, we concentrate on the region $0<s \lesssim 0.1$, which describes a finite-depth fractal glass, reminiscent of the Gardner phase (see Fig.~\ref{fig:zooming}). The evolution of the MSD upon varying $s$ is depicted in Fig.~\ref{fig:msdsmallbeta} and discloses many interesting traits: {\it i)} At short times ($0<t<\pi t_s$), there exists a universal regime, in which all curves collapse into a single one;
{\it ii)} The small plateaus regime sets in afterwards, but the overall slope of the intermediate-time behavior \textit{increases} as $s$  is reduced, thus showing a gradual transition from an overall glass to liquid phase;
{\it iii)} At sufficiently long times, there is saturation for all $s \neq 0$, but this freezing occurs on increasingly longer timescales as $s$ is reduced;
{\it iv)} These plateaus appear at fixed intervals in time, as promptly visualized in a linear scale plot (inset of Fig.~\ref{fig:msdsmallbeta}). Put together, all these features closely resemble (see SM) the marginal glass phase that has recently been observed in colloidal glass experiments~\cite{hammond2020experimental} and the emergent frequency that we obtain indicates that we are describing a \textit{Time Glass} phase. These results are further corroborated by calculations of the position and velocity auto-correlation functions, which show the same periodicity (see SM).

To find the coefficient of the time scale such that we have the period of the system, we concentrate on the case $s=0$ and put $v_0=0$ for simplicity. As this case is just a driven harmonic oscillator, we get
\begin{align}
\langle x(t)^2 \rangle &= \frac{K}{M^2} \int_0^t  \underbrace{\left[\tau E_{2,2}\left(-\frac{\eta}{M}     \tau^{2}\right) \right]^2}_{=\sin^2\left(\sqrt{\frac{\eta}{M}}\tau\right)}\;d\tau \nonumber\\
&= \frac{K}{2M\eta} \left[ t+ \frac{1}{2}\sqrt{\frac{M}{\eta}}\sin\left( 2\sqrt{\frac{\eta}{M}} t \right) \right],
\end{align}
where the relation between the Mittag-Leffler function and the sine can be seen using their Taylor expansions. This yields a periodicity of $\pi \sqrt{M/\eta}$ in time, thus providing the general period $P=\pi(M/\eta)^{1/(2-s)}=\pi t_s$ of the system. For non-zero $s$, this period applies only in a finite time-window before freezing.

\textbf{Microscopic Model.}
 Now, we aim at identifying an underlying microscopic model, which is phenomenologically described by the fractional Langevin equation. We will consider an open quantum system, upon which we perform a mean-field approximation to obtain an effective model. Since Hamiltonian dynamics relies on conservation of energy, we have to couple a system undergoing friction to a bath, which exerts that force. Inspired by a generalization of the Caldeira-Leggett model~\cite{caldeira1993dissipative,ferrer2007dynamical,caldeira1983path,caldeira1983tunneling,caldeira1985influence}, we consider a Hamiltonian
\begin{equation}
H= H_S +H_B +H_{int},
\end{equation}
where 
\begin{equation}
H_S= \frac{\hat{p}^2}{2M} +V(x)
\end{equation} 
is a Hamiltonian describing a ``Brownian'' particle, with mass $M$, momentum $p$, coordinate $x$ and subject to a potential $V(x)$;
\begin{equation}
 H_B= \sum_j \frac{\hbar \omega_j}{2} \sigma_{z j}
\end{equation}
describes a TLS bath (i.e. truncated harmonic oscillators) with natural frequencies $\omega_j$ and 
\begin{equation}
H_{int}= -x \sum_j J_j \sigma_{x j}
\end{equation}
is the interaction between the bath and the ``Brownian'' system~\cite{caldeira1993dissipative}. The degrees of freedom in the bath are then integrated out to describe quantum dissipation in the system. The spectral function is given by the imaginary part of the Fourier transform of the retarded dynamical susceptibility of the TLS bath, namely,
\begin{equation}
J(\omega):= \operatorname{Im} \mathcal{F}\left\langle-i \Theta(t-t') \left[F(t),F(t')\right] \right\rangle
\end{equation}
and it is crucial for connecting the microscopic parameters of the Hamiltonian with the phenomenological viscoelasticity coefficient $\eta$ appearing in the Langevin equation. Here, $F(t)$ is the force produced by the particle on the bath. For an Ohmic bath of harmonic oscillators, the spectral function is given by $J(\omega)=\eta \omega$ for $\omega<\Omega$, where $\Omega$ denotes a cutoff frequency, and is zero otherwise~\cite{caldeira2014introduction}. Such a bath will then be effectively described by the Langevin equation, where the friction is proportional to the velocity (first derivative of position) of the system. However, for our choice of sub-Ohmic TLS bath~\cite{ferrer2006optical,ferrer2007dynamical}
, the spectral function is given by 
\begin{equation}
J(\omega, T) = \eta\sin\left(\frac{\pi s}{2}\right) \omega^s \tanh\left(\frac{\hbar\omega}{2k T}\right) \Theta(\Omega-\omega),
\end{equation}
where $0<s<1$ and $\Omega$ is a cutoff frequency. The hyperbolic tangent is important for the quantum description where it allows the temperature to select which bath components are more relevant, namely, the ones with frequencies $\omega > 2k_B T/\hbar$. In our semi-classical approach, however, we take the limit $k T \ll \hbar \Omega$, which reduces it to the Caldeira-Leggett model, apart from losing equipartition. We can therefore rescale the TLS energies by $\omega^{1-s} t_s^{1-s}$ to retain a white-noise correlation. The friction force is then given by
\begin{equation}
F_{fr}= \frac{2}{\pi}\frac{d}{dt} \left\{  \int_0^t  \int_0^\infty \frac{J(\omega)}{\omega} \cos[\omega(t-t')]q(t') \text{ d}\omega\text{ d}t' \right\},
\end{equation}
which, after a reparametrization $\omega \to \omega/(t-t')$, becomes proportional to a fractional derivative in the limit $k T \ll \hbar \Omega$. The proportionality constant is then calculated by a complex contour integral, which results in a finite value for $0<s<1$ (see SM for details of the calculation). This then leads to a friction term given by
\begin{equation}
F_{fr} = \eta \mathbf{D}_{t}^s q(t).
\end{equation}
Therefore, a particle interacting with a sub-Ohmic two-level systems bath is well described by the fractional Langevin equation with white noise in the low-temperature limit.

\textbf{Conclusion.} 
 Making use of fractional Caputo derivatives, we have shown that a semi-classical system coupled to a low-temperature sub-Ohmic bath of two-level systems can be described by the fractional Langevin equation with white noise. We have solved this equation analytically and analyzed the anomalous diffusion.  
 Different behaviors were observed, depending on $s$, from ordinary Brownian motion for $s=1$ to glassy behavior for $s\lesssim 0.5$, a Time Glass for $0<s \lesssim 0.1$, and a marginal glass for $s=0$. Our work extends the use of fractional derivatives to the realm of sub-diffusion, by linking the formalism to the description of the Gardner transition. We identified a new regime between the Gardner phase and the usual glass, and showed that it is a realization of the long sought Time Glass. Further extensions of this model using the techniques applied to the fractional Langevin equation with colored noise~\cite{deng2009ergodic,pottier2003aging,metzler2014anomalous} are anticipated.


\begin{acknowledgments}
We are grateful to L.~M.~C. Janssen for introducing us to the Gardner phase and to P.~A. Zegeling for giving us insight into fractional calculus. We also thank J. de Graaf, S. Franz, M. Katsnelson, W. van Saarloos, J. Wettlaufer, T.~H. Hansson, E. Barkai, B. Bu\v{c}a, and F. Wilczek for fruitful discussions. R.~F.~O. is partially supported by Coordena\c{c}\~ao de Aperfei\c{c}oamento de Pessoal de N\'{\i}vel Superior (CAPES), finance code 001.
\end{acknowledgments}


\clearpage
\appendix


\section*{Supplementary material}

# Time Glass: A Fractional Calculus Approach
# Supplementary Material

R. C. Verstraten[1], R. F. Ozela[1,2], C. Morais Smith[1]

[1]*Institute for Theoretical Physics, Utrecht University,
Princetonplein 5, 3584CC Utrecht, The Netherlands*

[2]*Faculdade de Física, Universidade Federal do Pará, 66075-110 Belém, PA, Brazil*

(Dated: May 7, 2021)

## 1. FRACTIONAL CALCULUS: A BRIEF REVIEW

The first recorded mentioning of a fractional derivative was made by Leibniz in a letter to l'Hôpital in 1695, which motivated Euler in 1738 to introduce the Gamma function [1]. Fourier suggested a fractional derivative based on trigonometric functions in 1822, although the more complete foundations were only done by Liouville in 1832. An intriguing paradox arose, where the exponential function no longer was its own fractional derivative when expanded in its Taylor series. In 1847, Riemann found a more elegant way to derive Liouville's result, as a certain integral combined with an ordinary derivative, which became known as the Riemann-Liouville fractional derivative. Finally, in 1867, Grünwald understood that, by looking at the negative order $p = -1$ (i.e. an integral), there was a different base-point for some definitions, which lead to the conclusion that fractional derivatives could never be a local operator, and it was much more similar to an integral operator. More details about the history and development of fractional derivatives can be found in Ref. [1].

After the discovery of some applications and a new definition by Caputo in 1969 [2], the community paid new attention to the topic. It inspired books by Oldham and Spanier (1974) [3], Samko, Kilbas, and Marichev (1993) [4], Podlubny (1999) [5], and Hilfer (2000) [6], among others. Caputo only altered Riemann-Liouville's definition by interchanging the order of operation. However, for practical applications this made a huge difference, as this new definition was the first fractional derivative to maintain ordinary boundary conditions in a fractional differential equation. For details, we refer the reader to Ref. [7].

We will use the notation $^X_a\mathbf{D}_t f(t)$ for the fractional derivative named after Riemann-Liouville, Liouvile, Caputo, and Weyl, with notation $X \in \{RL, L, C, W\}$, respectively. Here, we show various properties of these fractional derivatives. For the rest of this section, let $p, q > 0$, $\nu \in \mathbb{R}$, $n-1 \leq p < n$, and $m-1 \leq q < m$, with $n, m \in \mathbb{N}$. The Riemann-Liouville fractional integral is based on a trick, where an $n^{th}$ order integral is simplified by reversing the integration order, leaving only one non-trivial integral. After generalizing the factorial to a Gamma function, we find the Riemann-Liouville fractional integral (denoted by a negative order), given by

$$^{RL}_a\mathbf{D}_t^{-p} f(t) = \frac{1}{\Gamma(p)} \int_a^t (t-\tau)^{p-1} f(\tau)\, d\tau. \quad (1)$$

The Riemann-Liouville fractional derivative is then taken by an integer derivative of this integral [8]

$$^{RL}_a\mathbf{D}_t^p f(t) = \frac{d^n}{dt^n}\, {}^{RL}_a\mathbf{D}_t^{p-n} f(t). \quad (2)$$

These Riemann-Liouville fractional operators act on powers as

$$^{RL}_a\mathbf{D}_t^{\pm p}(t-a)^\nu = \frac{\Gamma(\nu+1)}{\Gamma(\nu \mp p + 1)}(t-a)^{\nu \mp p}, \quad (3)$$

which means that the derivative of a constant may be non-zero. This is partially due to the additivity of orders being counter intuitive. For integer derivatives, derivative orders naturally add up, but for fractional derivatives, boundary terms start to appear, in the same sense as the fundamental theorem of calculus, where

$$\frac{d}{dt}\int_a^t f(\tau)\, d\tau = f(t), \quad (4)$$

$$\int_a^t f'(\tau)\, d\tau = f(t) - f(a). \quad (5)$$

For fractional derivatives, however, the additivity of orders are:

$$^{RL}_a\mathbf{D}_t^{\pm q}\, {}^{RL}_a\mathbf{D}_t^{-p} f(t) = {}^{RL}_a\mathbf{D}_t^{\pm q - p} f(t), \quad (6)$$

$$^{RL}_a\mathbf{D}_t^{-p}\, {}^{RL}_a\mathbf{D}_t^{q} f(t) = {}^{RL}_a\mathbf{D}_t^{q-p} f(t)$$
$$- \sum_{j=1}^{m} \left({}^{RL}_a\mathbf{D}_t^{q-j} f\right)(a) \frac{(t-a)^{p-j}}{\Gamma(p-j+1)}, \quad (7)$$

$$^{RL}_a\mathbf{D}_t^{p}\, {}^{RL}_a\mathbf{D}_t^{q} f(t) = {}^{RL}_a\mathbf{D}_t^{p+q} f(t)$$
$$+ \sum_{j=1}^{m} \left({}^{RL}_a\mathbf{D}_t^{q-j} f\right)(a) \frac{(t-a)^{-p-j}}{\Gamma(-p-j+1)}. \quad (8)$$

In particular, Eq. (8) suggests a difference from integer derivatives. However, since the Gamma function has poles exactly at the negative integers, this formula reduces to the well known additivity property if $p$ is an integer.

Laplace transforms, for instance, are useful for solving fractional differential equations, and for the Riemann-Liouville derivative they are given by

$$\mathcal{L}\left[{}^{RL}_{0}\mathbf{D}^{-p}_{t}f(t);s\right] = s^{-p}F(s), \quad (9)$$

$$\mathcal{L}\left[{}^{RL}_{0}\mathbf{D}^{p}_{t}f(t);s\right] = s^{p}F(s) - \sum_{k=0}^{n-1} s^{k}\left({}^{RL}_{0}\mathbf{D}^{p-k-1}_{t}f\right)(0). \quad (10)$$

Another convenient tool to work with fractional derivatives is the generalization of the exponential, called the Mittag-Leffler function

$$E_{\alpha,\beta}(z) = \sum_{k=0}^{\infty} \frac{z^k}{\Gamma(\alpha k + \beta)}, \quad (11)$$

where $\alpha, \beta > 0$, which reduces to the exponential if $\alpha = \beta = 1$.

The Caputo definition only holds for derivatives

$$^{C}_{a}\mathbf{D}^{p}_{t}f(t) = {}^{RL}_{a}\mathbf{D}^{p-n}_{t}f^{(n)}(t), \quad (12)$$

while its integral form can be seen as the regular Riemann-Liouville integral. Unlike Riemann-Liouville derivatives, the Caputo derivative of a constant will be zero. We used a simplified notation $\mathbf{D}^{p}_{t} = {}^{C}_{0}\mathbf{D}^{p}_{t}$ in the main text. An important property that one has to be careful with while changing the order of Caputo derivatives is

$$\lim_{\alpha \uparrow n} {}^{C}_{a}\mathbf{D}^{\alpha}_{t}f(t) = \lim_{\alpha \downarrow n} {}^{C}_{a}\mathbf{D}^{\alpha}_{t}f(t) + f^{(n)}(a) = f^{(n)}(t), \quad (13)$$

as it is discontinuous at the integers. A more comparable relation between the Riemann-Liouville and Caputo derivatives is given by

$$^{RL}_{a}\mathbf{D}^{p}_{t}f(t) = \sum_{k=0}^{n-1} \frac{f^{(k)}(a)(t-a)^{k-p}}{\Gamma(k-p+1)} + {}^{C}_{a}\mathbf{D}^{p}_{t}f(t). \quad (14)$$

This implies that the two definitions are equal if $f^{(k)}(a) = 0$ for all positive integers $k < p$, with $p$ the order of the derivative. The Laplace transform of the Caputo derivative is given by

$$\mathcal{L}\left[{}^{C}_{0}\mathbf{D}^{p}_{t}f(t);s\right] = s^{p}F(s) - \sum_{k=0}^{n-1} s^{p-k-1}f^{(k)}(0). \quad (15)$$

Remark the difference with Riemann-Liouville in Eq. (10); we essentially moved the fractional term from the order of the derivative to the power of $s$.

A particular case of the Riemann-Liouville definition is when $a = -\infty$, called the Liouville fractional derivative. From Eq. (8), a divergence might appear when $a \to -\infty$. The usual way to prevent this is by assuming that $f^{(n)}(t)t^{\alpha} \to 0$ as $t \to -\infty$, for all $n = 0, 1, \ldots, K$ and all $\alpha < K$, with $K$ larger than the order of the derivative.

One particular benefit of this choice is that it is compatible with Fourier transforms, serving as a bridge between Riemann-Liouville and Weyl fractional derivatives.

The Weyl definition is directly based on Fourier transforms

$$^{W}\mathbf{D}^{\pm p}_{t}f(t) = \mathcal{F}^{-1}\left[(i\omega)^{\pm p}\mathcal{F}\left(f(t);\omega\right);t\right], \quad (16)$$

which also implies that one needs the entire history of a function in order to use it. As long as a function satisfies both the requirements of the derivative definition and is Fourier transformable, any definition that is compatible with Fourier transforms, like the Liouville derivative, can be shown to be equivalent to the Weyl definition for those functions.

## 2. GLASS PHASES

In this section, we address some glass-related states in physics. Although a normal glass macroscopically resembles a crystal, their inner structures are nothing alike. The particles (atoms, molecules, or colloids) forming a crystal are periodically arranged, while they are randomly distributed in a glass. Glasses are not alike to liquids either, although they can flow on long time scales. Due to interactions with their neighbors, the particles in a glass can get trapped in effective cages. This prevents the particles from freely moving around, distinguishing a glass from a liquid. The liquid-to-glass transition is, therefore, often described by the friction coefficient $\eta$ diverging to such a high value that the "liquid" can be considered a solid on any relevant time scale [9]. In other words, the Mean Square Displacement (MSD) of a glass particle is notably similar to the movement of crystalline particles: ballistic movement up to a cage size where the MSD saturates [see Fig. 1(f)]; However, there is a major difference: the crystalline phase represents the absolute minimum of the energy landscape, while the amorphous cages of the glass are relatively higher metastable basins [10].

Many studies on glassy phases investigate the so-called glass-forming liquid in the supercooled region [10]. In this phase, the asymptotes of the MSD look like a liquid, but on intermediate time scales the MSD shows a plateau [11]. These systems can be understood as liquid phases with temporary *frozen intermediate states* associated to the metabasins characteristic of glasses. They manifest effective cages in different sites spread through the liquid, but particles can occasionally jump between cages and are therefore less restricted in their movement.

Langevin-type equations have been used before to describe atomic, polymeric and colloidal systems [12] and glassy dynamics [13]. The microscopic quantities are treated as stochastic variables, which together with a mean-field approximation can lead to many types of Langevin equations [14], often written in the form of the

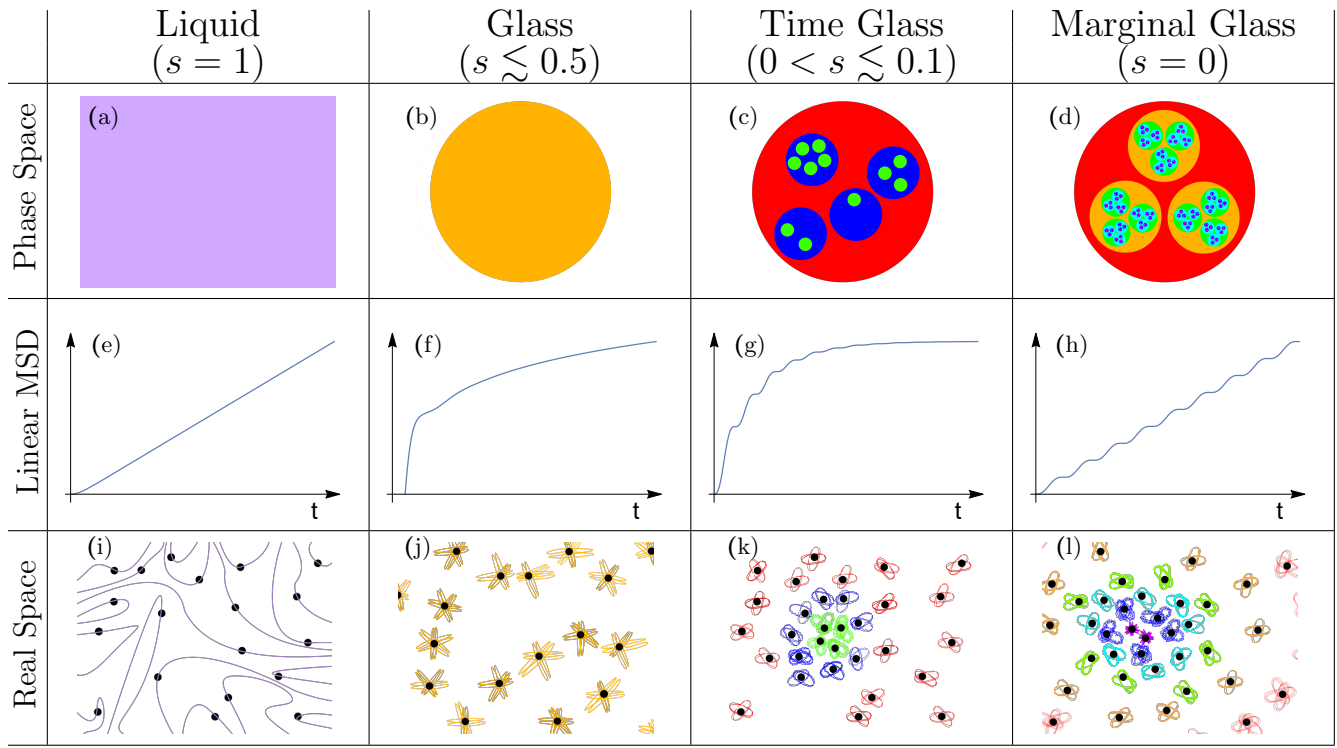

Fig. 1: Figure inspired by Ref. [17]. An overview of some phases that the fractional Langevin equation can predict. The phase space indicates all regions where a particle is allowed to go. The linear plot MSD shows the same trajectories as described in the main text, and are reprinted for clarity. The real space illustrates possible realizations of the particles movement. A Time Glass and a marginal glass have a dynamical picture, so different colors are used to indicate cases at subsequent times.

so-called generalized Langevin equation

$$M\frac{d^2}{dt^2}x(t) + \eta \int_0^t K(t-t')\dot{x}(t')\,dt' = f(t), \quad (17)$$

where $\langle f(t)f(t')\rangle = \eta k_B T K(|t-t'|)$ and $K(t)$ is the kernel used to match the friction behavior in the systems. Notice that the kernel $K(t) = 2\delta(t)$ gives the Langevin equation, while the power-law kernel $K(t) \sim t^{-\alpha}$ provides a fractional friction of the form $\eta \,{}^C_0\mathbf{D}^\alpha_t x(t)$ for $0 < \alpha < 1$ and a colored noise for the force correlation, thus describing the fractional Langevin equation. In Ref. [12], the (generalized) Langevin equation is derived by applying the Dyson decomposition to the normalized velocity of the particles. For an insight on the generation of plateaus in the MSD as an indicative of glass dynamics through Mode-coupling theory (MCT), we refer the reader to Ref. [15], where a generalized Langevin equation arises, with a friction kernel determined by a self-consistency equation involving the density matrix and propagators of the scattering functions.

The appearance of periodic plateaus in Fig. 2(c) and 2(d) of the main text is an indicative of the glass dynamics discussed above. The case of infinitely many evenly-spaced plateaus ($s = 0$) provides a marginally stable glass with a self-similar fractal inherent structure, while the finitely many plateaus ($0 < s \lesssim 0.1$) represent a finite-depth (i.e. a finite generation) fractal hierarchy, suggestive of a Time Glass.

A Time Glass has been previously discussed in Ref. [16, p.29] in the context of Many-Body Localized (MBL) systems. In order to contrast those properties with our model, we associate each plateau in our MSD for $0 < s \lesssim 0.1$ with a typical cage size $\Delta_n$ nested inside each other. If we focus on a single particle, a typical velocity $v_T$ can be associated to the temperature, resulting in many typical frequencies $v_T/2\Delta_n$ associated to the particle bouncing back and forth in the several cages inside one-another. The cages corresponding to the MSD are, however, indicating typical sizes that appear only on average; any individual particle in a simulation, for example, will be subject to a different noise, producing a different caging, and thus several local incommensurate frequencies. By the nature of the Langevin equation, our system also does not exhibit long-range spatial order and therefore satisfies the Time-Glass properties indicated in Ref. [16], meaning that we effectively provide a single-particle realization of such MBL systems. In addition, our system shows an emergent periodicity in time, which is not mentioned by Ref. [16].

A schematic overview of the four most prominent phases discussed in the main article is given in Fig. 1.

- The phase space (first row) represents the different

areas where particles can move to. In the case of a liquid, this is the entire space [Fig. 1(a)], but in normal glasses this is restricted to a single cage [Fig. 1(b)]. We find a finite number of cages nested inside each other for a Time Glass [Fig. 1(c)], while there is an infinite hierarchy for a marginal glass [Fig. 1(d)].

- The MSD's are recast (second row) from the main article to foster the analogy. All phases show ballistic motion before the particles have had time to interact with other particles. The motion converts to linear diffusion for a liquid [Fig. 1(e)], while for a normal glass the movement saturates at the cage size [Fig. 1(f)]. A varying number of plateaus appear before saturation in the Time Glass regime [Fig. 1(g)], while for a marginal glass there is an infinite number of plateaus without saturation [Fig. 1(h)].

- The real space representation (third row) shows the typical movement of particles in the system. Liquid particles are free to move around and occasionally bump into their neighbors [Fig. 1(i)]. In contrast, glass particles are not able to move past their neighbors and are stuck to oscillating in place [Fig. 1(j)]. A Time Glass is a little more subtle, since we draw a dynamical picture where different particles are confined by subsequent cage sizes, indicated by differing colors, in an overall glassy phase [Fig. 1(k)]. A marginal glass shares the same dynamical representation, but is singled out because it has infinitely-many different cages without a final saturation [Fig. 1(l)].

## 3. SOLUTION FRACTIONAL LANGEVIN

We will discuss the fractional Langevin equation, which is a generalization of the Langevin equation for Brownian motion, where the friction is changed to a more general fractional derivative (using $\beta$ instead of $s$ in this section to avoid confusion with the Laplace variable $s$):

$$M \frac{d^2}{dt^2} x(t) + \eta \, {}_0^C\mathbf{D}_t^\beta x(t) = f(t), \qquad (18)$$

with $M$ the mass of a particle, $\eta$ a type of viscoelasticity, $f(t)$ a white-noise force with statistics $\langle f(t) \rangle = 0$ and $\langle f(t)f(t') \rangle = K\delta(t-t')$, and $\beta > 0$ with $m-1 \leq \beta < m$ for an integer $m$. The Laplace transform of Eq. (18) gives

$$Ms^2 X(s) + \eta s^\beta X(s) - Msx(0) - Mx'(0)$$
$$- \eta \sum_{j=0}^{m-1} s^{\beta-j-1} x^{(j)}(0) = F(s), \qquad (19)$$

which can be rewritten into

$$X(s) = \frac{1}{Ms^2 + \eta s^\beta} \Bigg[ F(s) + Msx(0) + Mx'(0)$$
$$+ \eta \sum_{j=0}^{m-1} s^{\beta-j-1} x^{(j)}(0) \Bigg]. \qquad (20)$$

Since inverse Laplace transforms can be difficult to work with, it is often easier to identify solutions with Laplace transforms of common functions [8]. Therefore, we first rewrite Eq. (20) as

$$X(s) = \frac{F(s)}{M} \frac{s^{-\beta}}{s^{2-\beta} + \eta/M} + \frac{x(0)s^{1-\beta}}{s^{2-\beta} + \eta/M}$$
$$+ \frac{x'(0)s^{-\beta}}{s^{2-\beta} + \eta/M} + \sum_{j=0}^{m-1} \frac{\eta x^{(j)}(0)}{M} \frac{s^{-j-1}}{s^{2-\beta} + \eta/M}. \qquad (21)$$

Now, we use the Laplace transformation

$$\mathcal{L}[t^p; s] = \Gamma(p+1) s^{-(p+1)}, \qquad (22)$$

linearity in the Laplace transformation, and the power-series expansions

$$t^{q-1} E_{p,q}(at^p) = \sum_{k=0}^\infty \frac{a^k t^{pk+q-1}}{\Gamma(pk+q)}, \qquad (23)$$

$$\frac{s^{p-q}}{s^p - a} = \sum_{k=0}^\infty a^k s^{-kp-q}, \qquad (24)$$

to see that

$$\mathcal{L}[t^{q-1} E_{p,q}(at^p); s] = \frac{s^{p-q}}{s^p - a}, \qquad (25)$$

which is a simplified version of a result in Ref. [18]. With Eq. (25), we can now see that the inverse transform of $X(s)$ (replacing $\beta$ to $s$ again) is given by

$$x(t) = \frac{1}{M} \Bigg\{ f(t) * \left[ t E_{2-s,2}\left(-\frac{\eta}{M} t^{2-s}\right) \right] + Mx(0) \left[ E_{2-s,1}\left(-\frac{\eta}{M} t^{2-s}\right) \right]$$
$$+ Mx'(0) \left[ t E_{2-s,2}\left(-\frac{\eta}{M} t^{2-s}\right) \right] + \sum_{j=0}^{m-1} \eta x^{(j)}(0) \left[ t^{2-s+j} E_{2-s,3-s+j}\left(-\frac{\eta}{M} t^{2-s}\right) \right] \Bigg\}, \qquad (26)$$

where $f(t) * g(t) = \int_0^t f(t-\tau)g(\tau)\,d\tau$ is the convolution. In particular, if we choose $0 \leq s < 1$ and by symmetry set $x(0) = 0$, we get the solution

$$x(t) = \frac{1}{M} f(t) * \left[ t E_{2-s,2}\left(-\frac{\eta}{M} t^{2-s}\right) \right]$$
$$+ v_0 t E_{2-s,2}\left(-\frac{\eta}{M} t^{2-s}\right), \quad (27)$$

where we define $v_0 = x'(0)$. We can now apply the usual statistical mechanics tools to this solution, to find the Mean Squared Displacement (MSD),

$$\langle x(t)^2 \rangle = \left\langle \left\{ \frac{1}{M} f(t) * \left[ t E_{2-s,2}\left(-\frac{\eta}{M} t^{2-s}\right) \right] \right\}^2 \right\rangle$$
$$+ \left[ v_0 t E_{2-s,2}\left(-\frac{\eta}{M} t^{2-s}\right) \right]^2$$
$$+ 2 v_0 t E_{2-s,2}\left(-\frac{\eta}{M} t^{2-s}\right) \times$$
$$\frac{1}{M} \langle f(t) \rangle * \left[ t E_{2-s,2}\left(-\frac{\eta}{M} t^{2-s}\right) \right]$$
$$= \frac{K}{M^2} \int_0^t \left[ \tau E_{2-s,2}\left(-\frac{\eta}{M} \tau^{2-s}\right) \right]^2 d\tau$$
$$+ \left[ v_0 t E_{2-s,2}\left(-\frac{\eta}{M} t^{2-s}\right) \right]^2. \quad (28)$$

## 4. ASYMPTOTIC BEHAVIOR

We can expand Eq. (28) up to lowest non-zero order for $0 \leq t \ll (M/\eta)^{1/(2-s)}$. For this, we can simply use the lowest-order terms in the definition of the Mittag-Leffler function (Eq. 11) and insert these into the MSD to find

$$\langle x(t)^2 \rangle = \frac{K}{M^2} \int_0^t \left[ \tau \frac{1}{\Gamma(2)} \right]^2 d\tau + \left[ v_0 t \frac{1}{\Gamma(2)} \right]^2$$
$$= \frac{K}{3M^2} t^3 + v_0^2 t^2 = v_0^2 t^2 + \mathcal{O}(t^3), \quad (29)$$

which describes a freely moving particle. Indeed, this is expected, as on very short timescales there has not been enough time to meet a neighboring particle, and the motion should be ballistic.

On large timescales, we need some analysis on the asymptotic behavior of the Mittag-Leffler function. Lemma 1.1 in Ref. [19] provides that, for $p \in (0,2)$ and $q > 0$, we have

$$E_{p,q}(-z) = \frac{1}{z\Gamma(q-p)} + \mathcal{O}(|z|^{-2}) \quad \text{as } z \to \infty. \quad (30)$$

When inserting this relation into the MSD, we have to be careful with the integral term

$$\frac{K}{M^2} \int_0^t \left[ \tau E_{2-s,2}\left(-\frac{\eta}{M} \tau^{2-s}\right) \right]^2 d\tau, \quad (31)$$

as this still includes values from small time scales. Therefore, we split the integral into two separate parts, at a time $t_l$, such that we have a short timescale integral up to time $t_l$

$$MSD_{st} = \frac{K}{M^2} \int_0^{t_l} \left[ \tau E_{2-s,2}\left(-\frac{\eta}{M} \tau^{2-s}\right) \right]^2 d\tau \quad (32)$$

and an integral on long timescales $t > t_l$ where we can use the asymptotic relation

$$MSD_{lt}(t) = \frac{K}{M^2} \int_{t_l}^t \left[ \tau E_{2-s,2}\left(-\frac{\eta}{M} \tau^{2-s}\right) \right]^2 d\tau$$
$$= \frac{K}{M^2} \int_{t_l}^t \left[ \frac{\tau}{\frac{\eta}{M} \tau^{2-s} \Gamma(s)} \right]^2 d\tau$$
$$= \frac{K}{\eta^2 \Gamma(s)^2 (2s-1)} \left( t^{2s-1} - t_l^{2s-1} \right) \quad (33)$$

if $s \neq 0.5$ or

$$MSD_{lt}(t) = \frac{K}{\eta^2 \Gamma(0.5)^2} \log\left(\frac{t}{t_l}\right) \quad (34)$$

for $s = 0.5$. For the initial velocity term at $t > t_l$, we have

$$MSD_{v_0}(t) = \left[ v_0 t E_{2-s,2}\left(-\frac{\eta}{M} t^{2-s}\right) \right]^2$$
$$= \left[ v_0 t \frac{1}{\frac{\eta}{M} t^{2-s} \Gamma(s)} \right]^2$$
$$= \left( \frac{v_0 M}{\eta \Gamma(s)} \right)^2 t^{2s-2}. \quad (35)$$

We can now combine the separate parts to conclude that

$$\langle x(t \gg t_l)^2 \rangle = C(t_l) + \frac{K t^{2s-1}}{\eta^2 \Gamma(s)^2 (2s-1)} + \mathcal{O}(t^{2s-2})$$
$$\sim t^{2s-1} \quad (36)$$

if $s \neq 0.5$, where

$$C(t_l) = MSD_{st} - \frac{K t_l^{2s-1}}{\eta^2 \Gamma(s)^2 (2s-1)}, \quad (37)$$

and for $s = 0.5$, we find

$$\langle x(t \gg t_l)^2 \rangle \sim \log(t). \quad (38)$$

For $s \to 1$, we retrieve the familiar MSD of Brownian motion. However, a particularly important observation is that this long-time exponent in the MSD is negative for $0 < s < 0.5$, implying that the MSD saturates at a finite value after some time. This means that one goes through several regimes upon lowering $s$ from 1: for $s = 1$, there is a liquid state; then the movement becomes more and more restricted until $s = 0.5$, where one reaches a glassy state. Upon lowering $s$ further, there is a long term glassy state, but more plateaus emerge as the MSD starts to act more and more like a harmonic oscillator, while still saturating, until $s = 0$, where the harmonic oscillator behavior is completely retrieved, without any saturation.

## 5. NOISY UNDAMPED HARMONIC OSCILLATOR AS A MARGINAL GLASS

The solutions found in the $s = 0$ case for the fractional Langevin equation with white noise describe the same behavior that is known for a marginal glass. The fundamental mechanism causing this behavior is, however, an undamped harmonic oscillator driven by white noise. To explain how this mechanism can cause the properties of a marginal glass, we focus on an idealized path that a particle might take. In Fig. 2, we see the particle oscillating around the origin with a constantly increasing amplitude. Although the path will always be different and will depend on the specific realization of the white noise, we can argue that the particle will, on average, gain energy from this noise. For this, we focus on the total energy

$$E = \frac{1}{2}mv^2 + \frac{1}{2}\eta x^2 \qquad (39)$$

of the particle of interest, where the potential energy comes from the $s = 0$ effective friction term $\eta x$. When the particle is kicked, its velocity changes by $dv$. The energy of the particle after the kick is then given by

$$E' = \frac{1}{2}m(v + dv)^2 + \frac{1}{2}\eta x^2, \qquad (40)$$

which means that the change in energy from this kick is given by

$$\Delta E = E' - E = \frac{m}{2}\left(2v\,dv + dv^2\right). \qquad (41)$$

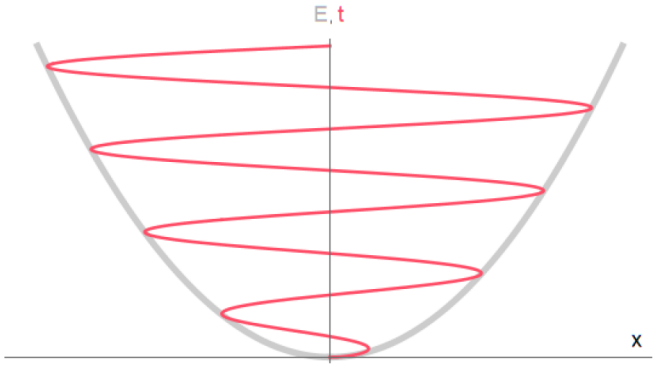

Fig. 2: Schematic drawing of an idealized path (red) in the harmonic potential (gray), which is characterizing the dissipative term for $s = 0$. As time increases, the particle gains energy on average, allowing it to go higher up the potential at each oscillation.

Now, we focus on the two different regions (growth and plateau) in Fig. 3: In the plateau region, the energy of the particle is dominated by the kinetic energy and $v$ is close to its maximal value. When a kick from the white noise occurs, this will change the velocity of the particle by a small amount $|dv| < |v|$ compared to the particle's velocity. Hence, the energy changes by $\Delta E \approx m\,v\,dv$. Since there is an equal chance that the kick is in the same or in the opposite direction as the particle velocity, $\Delta E$ is zero on average, leading to the plateau in the MSD. Now, we consider the growth regions, where $E$ is dominated by the potential energy and the velocity $v$ is small. For very small kicks (when $|dv| < |v|$), there is no gain in energy, but for strong kicks (when $|dv| > |v|$), we find that $\Delta E \approx \frac{m}{2}dv^2$, which indicates that both directions of the kick will increase the energy of the particle. Therefore, statistically, the particle will gain energy whenever it is near its maximal amplitude. Since the energy $E$ is directly related to the maximal amplitude, we find growth of the MSD in these regions, corresponding to the transition of the particle from a smaller to a larger cage. Hence, the plateaus occur with the same periodicity as induced by the harmonic potential.

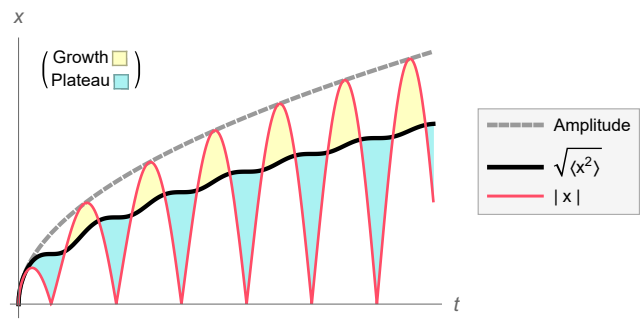

Fig. 3: Schematic drawing of the absolute value of the idealized path in Fig. 2 (red line), along with the square-root of the MSD (black line). Each time that the particle comes close to its maximal amplitude (dashed gray line), the MSD grows to the next plateau (yellow regions), while the plateaus occur when the particle is far away from its maximal amplitude (blue regions).

## 6. SUB-DIFFUSIVE QUANTUM HEAT BATH

The quantum description of an open system characterized by the Langevin equation in the semi-classical limit was proposed by Caldeira and Leggett [20–22]. They coupled the system of interest to a bath of harmonic oscillators and, by integrating out the reservoir, they have shown the influence of the bath on the dynamics of the particle. One important element in the Caldeira-Leggett description is the spectral function $J(\omega)$ of the bath, given in microscopic quantities by

$$J(\omega) := \operatorname{Im}\mathcal{F}\left\langle -i\Theta(t - t')\left[F(t), F(t')\right]\right\rangle, \qquad (42)$$

where $F(t)$ is the force exerted by the system on the bath, $\Theta$ is the Heaviside step function, and $\mathcal{F}$ denotes the Fourier transform.

Initially, the bath is described by microscopic variables such as the mass $m_j$, the frequency $\omega_j$, and spring constants $C_j$ of a set of harmonic oscillators, but rewriting

the spectral function, these microscopic quantities are connected to a phenomenological friction coefficient $\eta$ in the Langevin equation. Caldeira and Leggett consider systems described by Ohmic dissipation and, therefore, assume that the spectral function is linearly proportional to the frequency $\omega$. Later on, it was understood that a more generic bath behavior is possible, with the spectral function proportional to $\eta \omega^s$, where $0 < s < 1$ describes sub-Ohmic systems, such as proposed in Refs. [23, 24], while $s > 1$ accounts for super-Ohmic dynamics. Following the calculation from section 5.1 in Ref. [25], we find that the effective friction $F_{fr}$, in the Langevin equation, is given by

$$F_{fr} = \frac{2}{\pi} \frac{d}{dt} \left\{ \int_0^t \int_0^\infty \frac{J(\omega)}{\omega} \cos[\omega(t-t')] q(t') \, d\omega \, dt' \right\}. \tag{43}$$

Now, we consider a bath of two-level systems, such as the one introduced in Ref. [26], which has a spectral function of the form

$$J(\omega, T) = \eta \sin\left(\frac{\pi s}{2}\right) \omega^s \tanh\left(\frac{\hbar \omega}{2kT}\right) \Theta(\Omega - \omega), \tag{44}$$

with $\Omega$ a cutoff frequency. Although the hyperbolic tangent is not needed to produce the fractional Langevin equation in our semi-classical regime, it is the correct formalism to be adopted in studies which go beyond the semiclassical description in the low-temperature limit. The hyperbolic tangent allows for a selection of frequencies which could also be essential for studying phonons and fractons in this system. In the low-temperature limit, however, this will take the form

$$J(\omega) = \eta \sin\left(\frac{\pi s}{2}\right) \omega^s, \qquad \text{with } 0 < s < 1. \tag{45}$$

We now focus on the effective friction term $F_{fr}$ in the sub-Ohmic regime. When $t' = t$, we find a pole of order $s$, which is the same order found in a Caputo derivative of order $s$. For all other values of $t'$, we can make a change of variables $\omega \to \omega/(t-t')$ to get

$$\begin{aligned} F_{fr} &= \frac{2}{\pi} \eta \sin\left(\frac{\pi s}{2}\right) \frac{d}{dt} \left\{ \int_0^\infty \omega^{s-1} \cos(\omega) \, d\omega \int_0^t (t-t')^{-s} q(t') \, dt' \right\} \\ &= \frac{2}{\pi} \eta \sin\left(\frac{\pi s}{2}\right) \Gamma(1-s) \int_0^\infty \omega^{s-1} \cos(\omega) \, d\omega \, {}^{RL}_0\mathbf{D}_t^s q(t). \end{aligned} \tag{46}$$

We then observe that the assumption $q(0) = 0$ and Eq. (14) imply that

$${}^{RL}_0\mathbf{D}_t^s q(t) = {}^{C}_0\mathbf{D}_t^s q(t). \tag{47}$$

Now, we focus on the integral that is left in Eq. (46). Expanding the cosine into exponentials, we have

$$\begin{aligned} &\int_0^\infty \omega^{s-1} \cos(\omega) \, d\omega \\ &= \frac{1}{2} \int_0^\infty \omega^{s-1} e^{i\omega} \, d\omega + \frac{1}{2} \int_0^\infty \omega^{s-1} e^{-i\omega} \, d\omega \\ &= \frac{i^s}{2} \int_0^{-i\infty} \nu^{s-1} e^{-\nu} \, d\nu + \frac{i^{-s}}{2} \int_0^{i\infty} \nu^{s-1} e^{-\nu} \, d\nu, \end{aligned} \tag{48}$$

where we used two reparametrisations $\omega = \pm i\nu$.

From Eq. (48), we note that we can make two quarter circle complex contour integrations and combine this with the Cauchy integral theorem to conclude that both integrals are equal to the integral from $0$ to $\infty$, since $s < 1$. We therefore find that

$$\begin{aligned} \int_0^\infty \omega^{s-1} \cos(\omega) \, d\omega &= \frac{i^s + i^{-s}}{2} \int_0^\infty \nu^{s-1} e^{-\nu} \, d\nu \\ &= \cos\left(\frac{\pi s}{2}\right) \Gamma(s), \end{aligned} \tag{49}$$

and thus

$$F_{fr} = \frac{2}{\pi} \eta \sin\left(\frac{\pi s}{2}\right) \Gamma(1-s) \Gamma(s) \cos\left(\frac{\pi s}{2}\right) {}^{C}_0\mathbf{D}_t^s q(t). \tag{50}$$

We can further simplify Eq. (50) using Euler's reflection formula

$$\Gamma(s)\Gamma(1-s) = \frac{\pi}{\sin(\pi s)} \quad \forall s \notin \mathbb{Z}, \tag{51}$$

and the period doubling formula

$$\sin(\pi s) = 2 \sin\left(\frac{\pi s}{2}\right) \cos\left(\frac{\pi s}{2}\right), \tag{52}$$

such that we have

$$F_{fr} = \eta \, {}^{C}_0\mathbf{D}_t^s q(t). \tag{53}$$

## 7. WHITE NOISE

The change in spectral function also has consequences for the correlation of the noise term. The force is given in microscopic variables by

$$f(t) = \mathcal{L}^{-1}\left[\sum_j C_j \left(\frac{\dot{q}_j(0)}{s^2+\omega_j^2} + \frac{sq_j(0)}{s^2+\omega_j^2}\right); t\right] \quad (54)$$

$$= \sum_j C_j \left[q_j(0)\cos(\omega_j t) + \frac{\dot{q}_j(0)}{\omega_j}\sin(\omega_j t)\right] \quad (55)$$

Since equipartition breaks down at low temperatures, we may assume the correlations to be non-standard. We introduce the following microscopic correlations:

$$\langle q_j(0)\rangle = \langle \dot{q}_j(0)\rangle = \langle \dot{q}_j(0)q_{j'}(0)\rangle = 0, \quad (56)$$

$$\langle q_j(0)q_{j'}(0)\rangle = t_s^{1-s}\frac{k_B T}{m_j\omega_j^{s+1}}\delta_{jj'} = (t_s\omega_j)^{1-s}\frac{k_B T}{m_j\omega_j^2}\delta_{jj'}, \quad (57)$$

$$\langle \dot{q}_j(0)\dot{q}_{j'}(0)\rangle = t_s^{1-s}\frac{k_B T}{m_j\omega_j^{s-1}}\delta_{jj'} = (t_s\omega_j)^{1-s}\frac{k_B T}{m_j}\delta_{jj'}, \quad (58)$$

where $t_s$ is the typical time. Here, we have rescaled the energy of each oscillator $j$ by $(t_s\omega_j)^{1-s}$, without changing their frequency. In the semi-classical regime, this can be understood as an increase in amplitude. This rescaling can be seen as a temperature gradient across the bath when the oscillators are spatially distributed according to their frequency $\omega_j$. The average force is still zero, but the force squared correlation is then given by

$$\begin{aligned}\langle f(t)f(t')\rangle &= \left\langle \sum_j C_j\left[q_j(0)\cos(\omega_j t) + \frac{\dot{q}_j(0)}{\omega_j}\sin(\omega_j t)\right]\sum_{j'}C_{j'}\left[q_{j'}(0)\cos(\omega_{j'}t') + \frac{\dot{q}_{j'}(0)}{\omega_{j'}}\sin(\omega_{j'}t')\right]\right\rangle \\ &= \sum_{jj'}C_jC_{j'}\left[\langle q_j(0)q_{j'}(0)\rangle\cos(\omega_j t)\cos(\omega_{j'}t') + \frac{\langle \dot{q}_j(0)\dot{q}_{j'}(0)\rangle}{\omega_j\omega_{j'}}\sin(\omega_j t)\sin(\omega_{j'}t')\right. \\ &\quad \left. + \frac{\langle q_j(0)\dot{q}_{j'}(0)\rangle}{\omega_{j'}}\cos(\omega_j t)\sin(\omega_{j'}t') + \frac{\langle \dot{q}_j(0)q_{j'}(0)\rangle}{\omega_j}\sin(\omega_j t)\cos(\omega_{j'}t')\right] \\ &= \sum_{jj'}C_jC_{j'}\left[\frac{t_s^{1-s}k_BT}{m_j\omega_j^{s+1}}\delta_{jj'}\cos(\omega_j t)\cos(\omega_{j'}t') + \frac{t_s^{1-s}k_BT}{m_j\omega_j\omega_{j'}\omega_j^{s-1}}\delta_{jj'}\sin(\omega_j t)\sin(\omega_{j'}t')\right] \\ &= \sum_j C_j^2\frac{t_s^{1-s}k_BT}{m_j\omega_j^{s+1}}\left[\cos(\omega_j t)\cos(\omega_j t') + \sin(\omega_j t)\sin(\omega_j t')\right] \\ &= t_s^{1-s}k_BT\sum_j \frac{C_j^2}{m_j\omega_j^{s+1}}\cos[\omega_j(t-t')] \\ &= t_s^{1-s}k_BT\int_0^\infty d\omega \sum_j \frac{C_j^2}{m_j\omega_j^{s+1}}\delta(\omega-\omega_j)\cos[\omega_j(t-t')] \\ &= t_s^{1-s}k_BT\frac{2}{\pi}\int_0^\infty d\omega\frac{J(\omega)}{\omega^s}\cos[\omega(t-t')]. \end{aligned} \quad (59)$$

Since the low-temperature limit provides an effective $J(\omega) = \eta\sin\left(\frac{\pi s}{2}\right)\omega^s$, we can see that for any $s$ we are only left with an integral over the cosine, which provides the effective correlation

$$\langle f(t)f(t')\rangle = 2\sin\left(\frac{\pi s}{2}\right)t_s^{1-s}\eta k_B T\delta(t-t'). \quad (60)$$

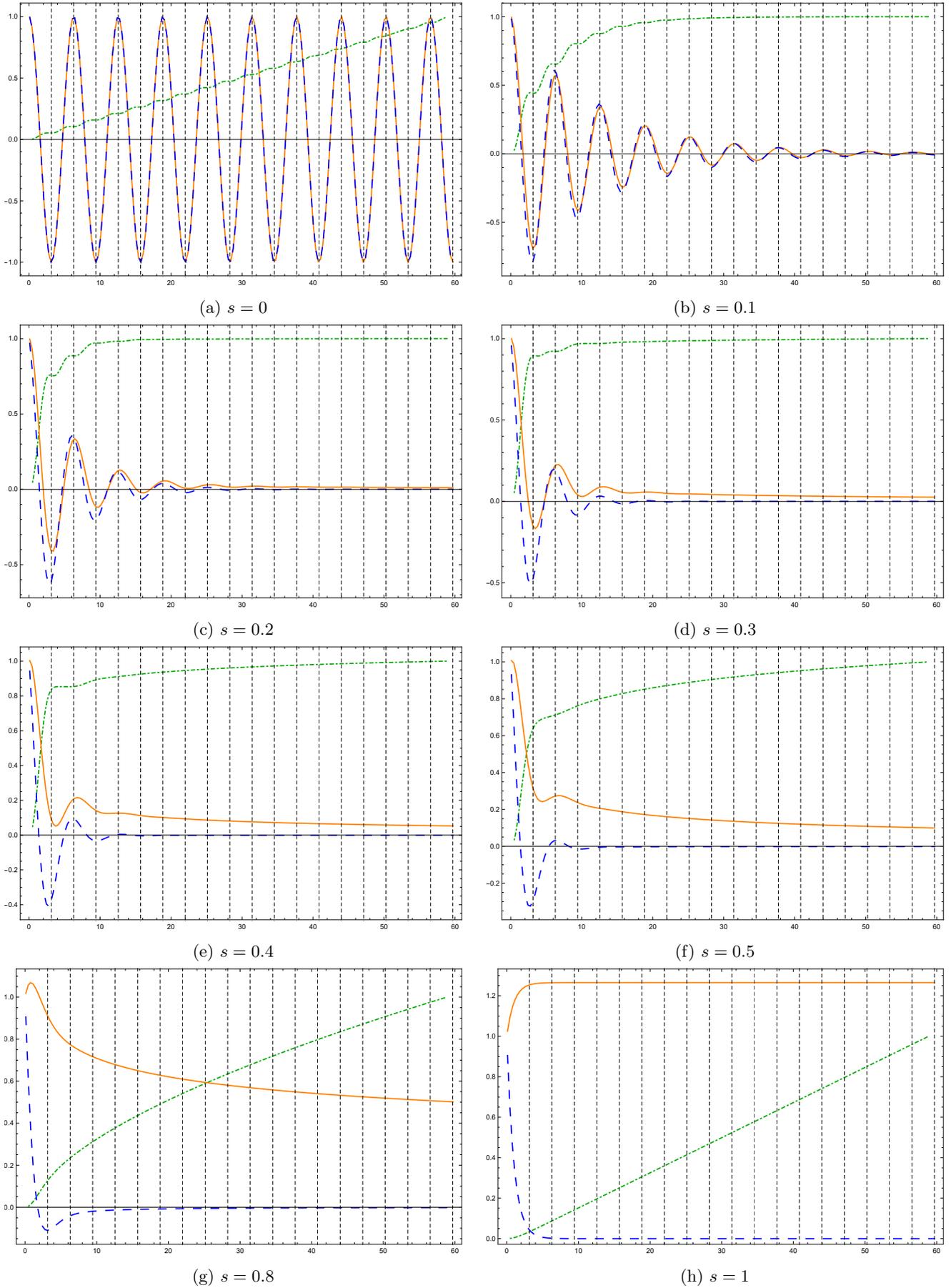

Fig. 4: Normalised MSD (Green, Dotdashed), PACF (Orange, Line), VACF (Blue, Dashed). The dashed vertical lines are at multiples of the emergent periodicity $\pi(M/\eta)^{1/(2-s)}$ and $M/\eta = 1$.

## 8. AUTOCORRELATIONS

The position and velocity autocorrelation functions (PACF and VACF, respectively) have been plotted alongside a normalized MSD for several values of $s$ in Fig. 4. Here, we can observe a clear relation between the plateaus in the MSD and the oscillations in the PACF and VACF. Upon lowering $s$ from one, we see small oscillations forming for a short initial period. These oscillations then become larger and remain for longer times, until at $s = 0$ they become a sine function in the harmonic oscillator. We want to highlight the striking similarity in the PACF with Ref. [27], even though their system is different with a colored noise and external harmonic potential. The analytical forms of the PACF and VACF for $t_0 \gg (M/\eta)^{s-2}$ are given by

$$\frac{\langle x(t_0)\, x(t_0+t)\rangle}{\langle x(t_0)^2\rangle} = \frac{K}{M^2\langle x(t_0)^2\rangle} \int_0^{t_0} \tau E_{2-s,2}\left(-\frac{\eta}{M}\tau^{2-s}\right)(t+\tau)E_{2-s,2}\left(-\frac{\eta}{M}(t+\tau)^{2-s}\right)\, \mathrm{d}\tau \quad (61)$$

$$\frac{\langle v(t_0)\, v(t_0+t)\rangle}{\langle v(t_0)^2\rangle} = \frac{K}{M^2\langle v(t_0)^2\rangle} \int_0^{t_0} \left[\frac{\mathrm{d}}{\mathrm{d}\tau}\tau E_{2-s,2}\left(-\frac{\eta}{M}\tau^{2-s}\right)\right]\left[\frac{\mathrm{d}}{\mathrm{d}(t+\tau)}(t+\tau)E_{2-s,2}\left(-\frac{\eta}{M}(t+\tau)^{2-s}\right)\right]\, \mathrm{d}\tau. \quad (62)$$


\let\clearpage\relax

\end{document}